# Ideal charge density wave order in the high-field state of superconducting YBCO

H. Jang[1,*], W.-S. Lee[2,*], H. Nojiri[3], S. Matsuzawa[3], H. Yasumura[3], L. Nie[4], A. V. Maharaj[4], S. Gerber[5], Y. Liu[1], A. Mehta[1], D. A. Bonn[6,7], R. Liang[6,7], W. N. Hardy[6,7], C. A. Burns[1,8], Z. Islam[9], S. Song[10], J. Hastings[10], T. P. Devereaux[2], Z.-X. Shen[2,4], S. A. Kivelson[4], C.-C. Kao[11], D. Zhu[10,†], J.-S. Lee[1,*,†]

[1]*Stanford Synchrotron Radiation Lightsource, SLAC National Accelerator Laboratory, Menlo Park, California 94025, USA*

[2]*Stanford Institute for Materials and Energy Science, SLAC National Accelerator Laboratory and Stanford University, Menlo Park, California 94025, USA*

[3]*Institute for Materials Research, Tohoku University, Katahira 2-1-1, Sendai, 980-8577, Japan*

[4]*Geballe Laboratory for Advanced Materials, Departments of Physics and Applied Physics, Stanford University, Stanford, California 94305, USA*

[5]*SwissFEL, Paul Scherrer Institut, 5232 Villigen PSI, Switzerland*

[6]*Department of Physics & Astronomy, University of British Columbia, Vancouver, British Columbia, Canada V6T 1Z1*

[7]*Canadian Institute for Advanced Research, Toronto, Ontario, Canada M5G 1Z8*

[8]*Dept. of Physics, Western Michigan University, Kalamazoo, Michigan, 49008, USA*

[9]*The Advanced Photon Source, Argonne National Laboratory, Argonne, Illinois 60439, USA*

[10]*Linac Coherent Light Source, SLAC National Accelerator Laboratory, Menlo Park, California 94025, USA*

[11]*SLAC National Accelerator Laboratory, Menlo Park, California 94025, USA*

*These authors contributed equally to this work.

†Correspondence to: dlzhu@slac.stanford.edu and jslee@slac.stanford.edu

The existence of charge density wave (CDW) correlations in cuprate superconductors has now been established [1-16]. However, the nature of the ground state order has remained uncertain because disorder and the presence of superconductivity typically limit the CDW correlation lengths to a dozen unit cells or less [7-20]. Here we explore the CDW correlations in $YBa_2Cu_3O_x$ (YBCO) ortho-II and ortho-VIII crystals, which belong to the cleanest available cuprate family [17], at magnetic fields in excess of the resistive upper critical field ($H_{c2}$) where the superconductivity is heavily suppressed [21].  We find an incommensurate, unidirectional CDW with a well-defined onset at a critical field strength that is proportional to $H_{c2}$. It is related to but distinct from the short-range bidirectional CDW that exists at zero magnetic field. The unidirectional CDW possesses a long in-plane correlation length as well as significant correlations between neighboring $CuO_2$ planes, yielding a correlation volume that is at least 2 - 3 orders of magnitude larger than that of the zero-field CDW. This is by far the largest CDW correlation volume observed in any cuprate crystal and so is presumably representative of the high-field ground-state of an "ideal" disorder-free cuprate.

Common characteristics of the CDW order in superconducting cuprates at zero magnetic field are bidirectional, quasi-two dimensional (2D) and short-range correlated [8,12-15]. More specifically, the CDW diffraction patterns are found in both directions of Cu-O bonds in the $CuO_2$ plane (Fig. 1a), and the CDW correlation lengths parallel and perpendicular to the planes (*i.e.,* along the *a*- or *b*-axes and the *c*-axis) are less than ~20 and ~1 lattice constants, respectively [7-16], corresponding to a correlation volume of order $10^2$ unit cells (UCs). Thus, the properties of the quasi-2D CDW are likely strongly affected by disorder, and only indirectly represent the true nature of the underlying CDW correlations. Indeed, x-ray scattering shows that the onset of the quasi-2D order is gradual without a sharp transition [7-16], consistent with the influence of quenched disorder on an incommensurate CDW [18,19]. Furthermore, while Y-based and La-based cuprates exhibit a clear competition between CDW and superconductivity [7,8,12-16], such competition is not apparent in the families of Bi-based and Hg-based cuprate-compounds [9-11] – a discrepancy that probably reflects different degrees of quenched disorder among cuprate families.

Recently, a CDW with significantly longer correlation lengths was observed in superconducting YBCO (Fig. 1b) via x-ray scattering at high magnetic fields [13, 14]. This reveals the character (*i.e.*, three-dimensional, 3D) of the high-field charge ordering previously inferred by other measurements [3-6]. At a magnetic field of ~ 30 Tesla, its in- and out-of-plane correlation lengths are of the order of 100 and 10 lattice constants, respectively [13,14], which are significantly larger than those of the zero field 2D CDW. Thus, it arguably represents the CDW ground state of an "ideal" disorder-free cuprate superconductor. However, to date, while this 3D CDW has been reported at doping levels of $p \sim 0.12$ and $\sim 0.11$ [13,14], limited high-field data near $H_{c2}$ has only been available at $p \sim 0.12$ [13]. To establish the 3D CDW phenomenology, it is important to track its doping and magnetic field dependences up to $H_{c2}$, and to elucidate its puzzling relationship with the quasi-2D CDW which coexists with the 3D one even at high-field (Fig. 1b) [13, 14].

In order to address these issues, we first investigate detwinned YBCO ortho-II ($x$ = 6.51) using x-ray scattering at an x-ray free electron laser (FEL) combined with a pulsed magnet (see Methods) [13]. With $\mu_0H = 28.2$ T, we clearly observe a field-induced CDW at $\boldsymbol{Q} = (0, 2-q, 1)$ with an incommensurate $q \sim 0.33$ r.l.u. (Fig. 1c). Similar to the case of YBCO ortho-VIII [13], the field-induced 3D CDW exhibits the same $q$-value as that of the zero-field quasi-2D CDW but with an integer rather than a half integer $l$-value [8, 15]. Note that the quasi-2D CDW still exists at this field (Extended Data Fig. 1b), confirming its coexistence with the 3D CDW.

The emergence of the 3D CDW order in YBCO ortho-II as a function of magnetic field is shown in Fig. 2a. As shown in the peak intensity at $l \sim 1$ (Fig. 2b), the 3D CDW appears at $\mu_0H_{3D} \sim 18$ T. Further increase of the magnetic field not only increases the diffraction peak intensity, but also narrows the peak width along both the $l$- and $k$-directions (Fig. 2c). At 28.2 T and 10K, the ortho-II correlation lengths in the $k$- and $l$-directions are more than 230 Å (which is limited by our instrument resolution) and 55 Å, respectively. Furthermore, the 3D CDW shows strong temperature dependence (Fig. 2d). At $\mu_0H = 26.0$ T, the projected peak intensity (Fig. 2e) and correlation length (Fig. 2f) demonstrated the appearance of the 3D CDW at $T_{3D} \sim 45$ K, which is considerably lower than the onset temperature of the quasi-2D CDW ($\sim 120$ K), but just slightly lower than $T_c$ (57 K at 0 Tesla). Note that the onset field and temperature agree well with the thermodynamic phase boundary deduced from sound velocity measurements [5]. Altogether, the $H$- and $T$-dependences demonstrate that 3D CDW emergence is triggered by the magnetic field suppression of superconducting correlations.

The field dependence of the 3D CDW in YBCO ortho-II is similar to that previously observed in ortho-VIII [13]. As shown in Fig. 3a, if plotted as a function of shifted magnetic field (*i.e.*, $\mu_0 H - \mu_0 H_{3D}$), the growth rate of the normalized 3D CDW intensity is remarkably similar in the two crystals, despite the different doping concentrations. Furthermore, both crystals exhibit a similar quantitative evolution of the 3D CDW correlation volume, which reaches $\sim 10^5$ UCs at $H \sim H_{c2}$ (Fig. 3b) – more than two orders larger than that of the quasi-2D order. Interestingly, the ratio, $H_{3D}/H_{c2}$ is approximately 0.6 for both crystals, suggesting that $H_{3D}$ closely tracks $H_{c2}$, a key characteristic of the superconducting state. These findings further demonstrate the intimate relation between the 3D CDW and superconductivity, as well as supporting the attribution of the 3D CDW order as representative of the "disorder-free" situation.

Finally, we investigate signatures of the 3D CDW at an ordering wavevector in the *h*-direction at $\sim H_{c2}$ [21]. In ortho-II, no sign of the CDW pattern is seen near *l =1* up to 30 T (Fig. 4a). This is also evident in the featureless projected intensity along the *l*-direction at $\boldsymbol{Q} = (1.67, 0, 1)$, in stark contrast to that at $\boldsymbol{Q} = (0, 1.67, 1)$. We have performed the same measurement on ortho-VIII (Fig. 4b), and similarly found no 3D CDW up to 25T, consistent with previous measurements up to 16.9 T [14]. These ordering absences for both ortho-II and ortho-VIII at $H \sim H_{c2} >> H_{3D}$ shows that the field-induced 3D CDW in YBCO is robustly unidirectional; the 3D CDW is "stripe-like" with an ordering vector parallel to the Cu-O chain direction.

We now discuss the puzzling relation between the unidirectional 3D CDW and the bidirectional 2D CDW. The drastic differences in the qualitative behaviors, including their directionality [13-16], dimensionality [8,13-15], and their *H*- and *T*-dependences [8,13-16], suggest that the unidirectional 3D CDW is a different entity than the quasi-2D CDW [7,8,15,16]. However, the diffraction signals from the 2D and 3D CDWs coexist [13, 14]. It seems to us that the most promising way to reconcile these observations is to assume that there are distinct domains with the two types of CDW. In this sense, the fact that the in-plane wavevectors of the two CDWs are identical, suggesting that they share the same local correlations inherent in the electronic structure of YBCO. Since the 3D CDW is unidirectional, we argue that the inherent CDW correlations correspond to unidirectional charge stripes.

On theoretical grounds [18], an incommensurate CDW phase exists as a sharply defined phase of matter (*i.e.,* with true long-range order) only in the ideal limit of vanishing disorder. However, as shown in Fig. 5, for a unidirectional CDW in a tetragonal system, a sharply defined nematic phase, a form of "vestigial" CDW order [18] that spontaneously breaks the point-group symmetry, exists so long as the disorder strength, $\sigma$, is less than a finite critical value, $\sigma_c$. The phase transition at $\sigma = \sigma_c$ is rounded in an orthorhombic system, but so long as the symmetry breaking field is weak, there remains a sharp crossover from an approximately bidirectional phase (*i.e.,* isotropic) for $\sigma > \sigma_c$ to a strongly unidirectional phase for $\sigma < \sigma_c$. The bidirectional phase can still be locally stripe-like, but with the orientation of the stripes determined by the local disorder potential rather than by the orthorhombicity. Indeed, a strong tendency to nematic order (oriented by the weak orthorhombicity of YBCO) has been inferred from various experiments [22-24]. In the Supplementary Information (see also the Extended Data Fig. 3), we illustrate these points using an effective field theory and also address the crossover from 2D to 3D correlations at $\sigma_c$. This leads us to interpret our results as suggestive of a universal tendency toward unidirectional incommensurate CDW order in YBCO and a somewhat non-uniform distribution of the disorder strengths. As shown in Fig. 5, upon the application of field, the isotropic 2D CDW in the less disordered region ($\sigma_{Less}$), transforms into the nematic 3D CDW phase, while that in the more disorder region ($\sigma_{More}$) still remains in the isotropic phase. This conjecture is consistent with the field-induced nematicity of the CDW near vortex cores hinted in a recent STM study on the double layer Bi-based cuprate [26], in which the influence of disorder is stronger than in YBCO.

In the context of other high-field experiments, note that quantum oscillations have been observed [27, 28] in ortho-II crystals (with $T_c \sim 58$–$60 K$) in fields above 18–22 T, $\sim H_{3D}$. Moreover, $H_{3D}$ agrees with the Fermi surface reconstruction field deduced from Hall coefficient measurements [29]. Although the proposed inhomogeneity picture qualitatively captures our experimental observations, it is not obvious that the NMR Cu and O lines are readily interpreted as the sum of contributions from a unidirectional and a bidirectional CDW [4, 6, 19].

This worry aside, our results strongly suggest that the ground state competing order in "ideal" superconducting YBCO with zero disorder would be a long range ordered, incommensurate, unidirectional CDW. Our results also lend support to the existence of nematic components in the proximate phases to the 3D CDW in the phase diagram, and to their interpretation as arising from remnant unidirectional 3D CDW correlations [18,30].

**Acknowledgments**: This work was supported by the Department of Energy (DOE), Office of Science, Basic Energy Sciences, Materials Sciences and Engineering Division, under contract DE-AC02-76SF00515. X-ray FEL studies were carried out at the Linac Coherent Light Source, a Directorate of SLAC and an Office of Science User Facility operated for the U.S. DOE, Office of Science by Stanford University. Resonant soft x-ray scattering measurements were carried out at the Stanford Synchrotron Radiation Lightsource (BL13-3), a Directorate of SLAC and an Office of Science User Facility operated for the U.S. DOE, Office of Science by Stanford University. H.N. acknowledges the support by KAKENHI 23224009, 15K13510, ICC-IMR and MD-program. C.A.B. was supported by the U.S. Department of Energy, Office of Basic Energy Sciences, Division of Materials Sciences and Engineering, under Award DE-FG02-99ER45772. Materials development was supported by the Natural Sciences and Engineering Research Council, and by the Canadian Institute for Advanced Research.

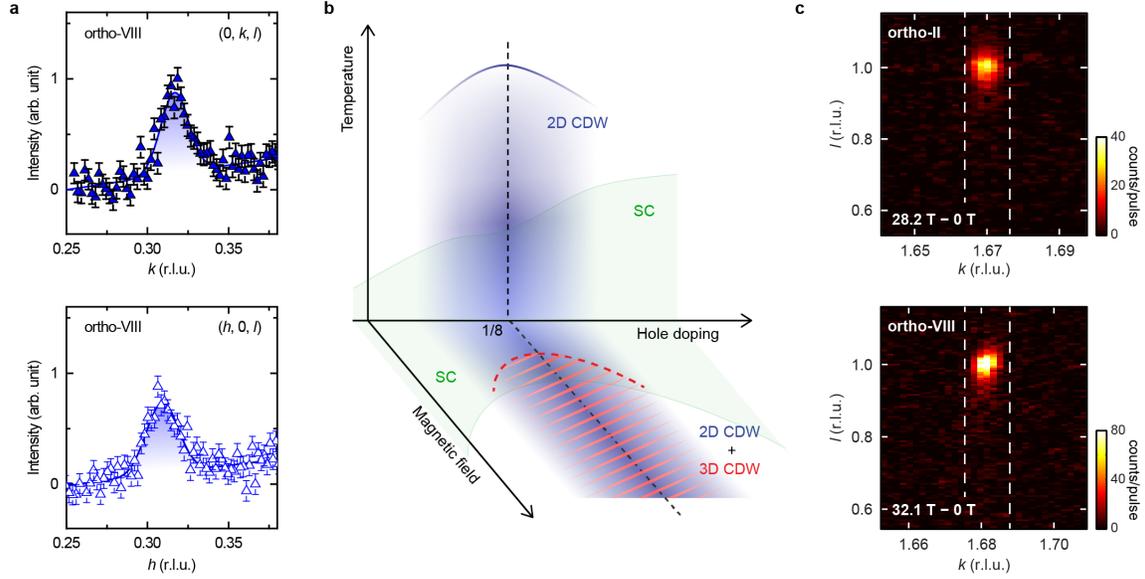

**Figure 1| Charge density wave orders in YBCO. a,** Zero-field quasi-2D CDW diffraction peak profiles (upper and lower panels) of ortho-VIII in the $k$- and $h$-directions, respectively, measured by resonant soft x-ray scattering at Cu $L_3$-edge and $T = 67$ K (see Method). It is bidirectional, *i.e.*, the diffraction peaks are found in both Cu-O bond directions in the $CuO_2$ plane. **b**, A schematic sketch of the temperature – doping – magnetic field phase diagram of YBCO near a hole concentration $p = 1/8$. The quasi-2D CDW (blue shaded area) evolves from high temperature and exists at both zero and finite magnetic fields. The 3D CDW (red colored, hatched area) emerges only in high magnetic fields and at low temperatures, but coexists with the 2D one (see also Extended Data Fig. 1b). **c,** Maps of the difference between the diffraction intensities at high-field and zero-field in ortho-II (upper panel) and ortho-VIII (lower panel) (See Extended Data Fig. 1a). The bright spots are the 3D CDW diffraction patterns located at $\boldsymbol{Q} \sim (0, 2\text{-}q, 1)$, where $q$ is ~0.33 in reciprocal lattice units (r.l.u.) and ~0.32 r.l.u. for ortho-II and ortho-VIII, respectively. The white dashed lines indicate the windows used to extract the projected peak profile shown in Figs. 2 and 4.

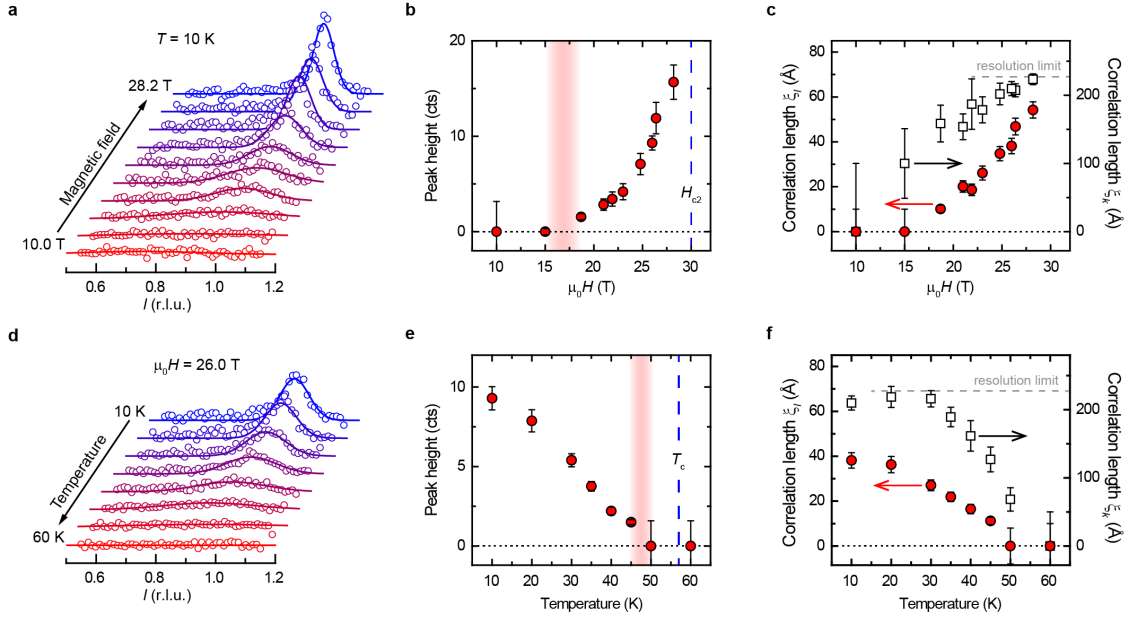

**Figure 2| Field and temperature dependences of the 3D CDW in ortho-II. a, d,** Projected peak profile along (0, 1.67, *l*) as a function of magnetic field at *T* = 10 K (**a**) and as a function of temperature at μ₀*H* = 26.0 T (**d**). The data points are obtained by integrating the field-induced signal over the range of *k* indicated by the dashed line in Fig. 1c. Solid lines are Gaussian fits to the data. Note that the *k*-*l* difference maps in the corresponding *H* and *T* are shown in Extended Data Fig. 2. **b, c,** Fitted 3D CDW peak heights (**b**) and correlation lengths (**c**) in the *l*- and *k*-directions as a function of magnetic field. **e, f,** Fitted 3D CDW peak heights (**e**) and correlation lengths (**f**) in the *l*- and *k*-directions as a function of temperature. The red shaded area denotes the onset region of the 3D CDW. Note, the *k*-correlation lengths at large *H* and low *T* are resolution limited (~ 230 Å indicated by the grey dashed lines in **e**, and **f**); thus, they represent lower bounds of the actual values. The displayed $\xi_k$ have been not made a resolution correction. All dotted lines indicate zero. The error bars denote 1 standard deviation (SD) as obtained from the peak fitting.

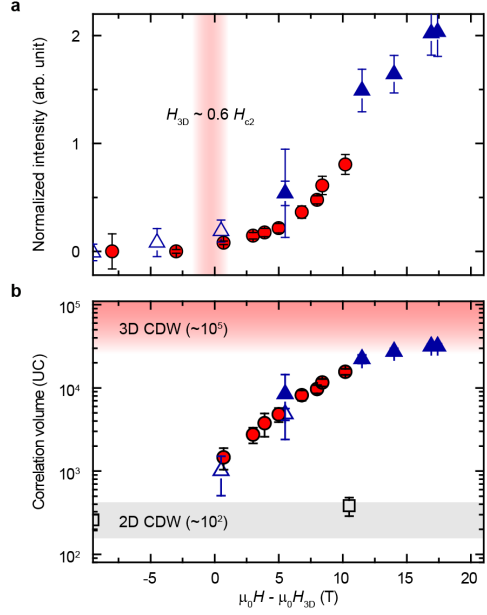

**Figure 3| Comparison of 3D CDWs in the ortho-II and ortho-VIII. a,** Fitted 3D CDW peak height of ortho-II (circles) and ortho-VIII (triangles) as a function of $\mu_0 H$ - $\mu_0 H_{3D}$. Here, $H_{3D}$ is 14.5 T and 18 T for the ortho-VIII [13, 14] and ortho-II crystals, respectively. The peak heights are normalized to 1 at $H_{c2}$. The values of resistive $H_{c2}$ are adapted from Ref. [21] – 24 T and 30 T for our ortho-VIII and ortho-II crystals, respectively. The red shaded area denotes the onset region of the 3D CDW. Data of open triangles were taken from Ref. [13]. **b,** Estimated 3D CDW's correlation volumes of ortho-II (circles) and ortho-VIII (open/closed triangles) as a function of scaled magnetic field. In order to estimate the correlation volume, we have assumed that the correlation length along the *h*-direction is the same as along the *k*-direction for the 3D CDW at (0, 2-*q*, 1). Note that the in-plane correlation length used in this estimate is resolution limited, so the estimated volume is a lower bound. The grey shaded area denotes the 2D CDW volume in ortho-VIII; data marked by the open squares (*i.e.,* 2D CDW volumes) were taken from Ref. [13]. Error bars correspond to 1 SD.

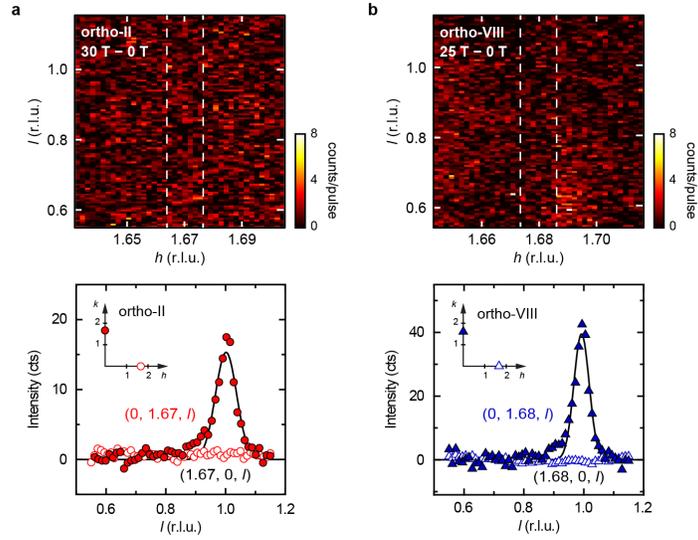

**Figure 4| Unidirectional character of the 3D CDW: a, b,** Zero-field-background-subtracted diffraction intensity maps of ortho-II (**a**) and ortho-VIII (**b**) in high magnetic fields and in the *hl*-reciprocal plane (upper panel). The lower panels display the projected intensities along $(1.67, 0, l)$ (red open circles) and $(0, 1.67, l)$ (red closed circles) in ortho-II and $(1.68, 0, l)$ (blue open triangles) and $(0, 1.68, l)$ (blue closed triangles) in ortho-VIII. The projected intensity is obtained by integrating the signal within the window indicated by the dashed lines in upper panels and Fig. 1c. Solid lines are Gaussian fits to the data.

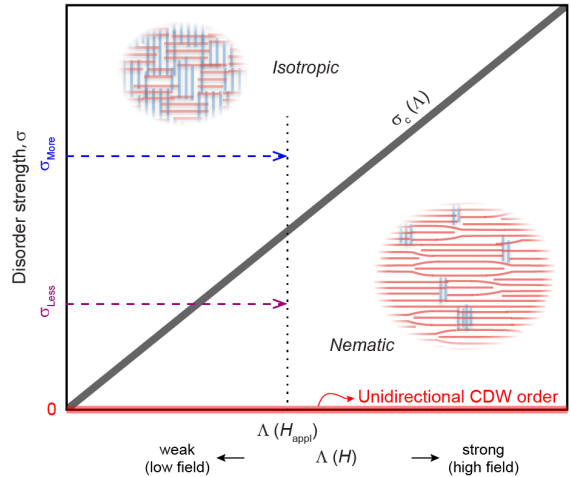

**Figure 5| Schematic CDW phase diagram.** Low temperature phase diagram of a layered crystal as a function of disorder strength ($\sigma$) and CDW strength, $\Lambda(H)$, assumed to be an increasing function of increasing $H$, due to the suppression of the superconductivity. It is assumed that the disorder-free state (marked by the red colored bar on the $x$-axis) is an incommensurate, unidirectional CDW. In a tetragonal crystal, the thick grey colored line marks a nematic transition, $\sigma_c$, while in a weakly orthorhombic YBCO crystal – it is a crossover. Above $\sigma_c$, CDW correlations are short ranged and bidirectional. Insets show the cartoons of disorder-pinned CDW domains; approximately isotropic, bidirectional, CDW phase (left top) and sharply defined nematic, unidirectional phase (right bottom). In the context of the proposed inhomogeneity scenario, the purple and blue dashed arrows demonstrate the field-dependent CDW evolution in the less ($\sigma_{Less}$) and more ($\sigma_{More}$) disordered regions of the sample. At a given applied magnetic field ($H_{appl}$), the $\sigma_{Less}$ region transforms from the bidirectional 2D CDW into the unidirectional 3D order, while the $\sigma_{More}$ region remains in the isotropic phase (see also the Extended Data Fig. 3).

**METHODS**

**Samples.** Detwinned single crystals of $YBa_2Cu_3O_{6.51}$ (ortho-II, $T_c = 57$ K, doping concentration $p \sim 0.1$) and $YBa_2Cu_3O_{6.67}$ (ortho-VIII, $T_c = 67$ K, $p \sim 0.118$) were studied. The single crystals were obtained from flux growth [31]. For two reflection-geometries as shown in the Extended Data Fig. 4, two crystals were prepared for each doping. The single crystals were parallel to the crystallographic $a$-axis and $b$-axis in $(0, k, l)$ and $(h, 0, l)$ reflection geometry, respectively, whilst applying the magnetic field $H$ along the c-axis. Note that we prepared thin crystals (less than 0.5 mm) along the $b$- and $a$-axis to avoid sample heating via eddy currents due the pulsed magnetic field.

**Resonant soft x-ray scattering under zero magnetic field.** The quasi-2D CDW from these crystals were characterized by resonant soft x-ray scattering (RSXS) measurements at the Cu $L_3$-edge (931 eV; maximum energy position of x-ray absorption spectroscopy) at the beamline 13-3 of the Stanford Synchrotron Radiation Lightsource (SSRL) as shown in Fig. 1a. Note that the obtained data were from theta (sample) scans at fixed detector angle 176°. An $l$-value of ~1.45 r.l.u., *i.e.*, slightly less than the half-integer value, had to be chosen due to experimental constraints and the limited total momentum transfer at $E = 931$ eV. All RSXS data background are subtracted by references that were measured at 150 K. Solid lines are Gaussian fits to the data with a linear background.

**Pulsed high magnetic field x-ray scattering setup at LCLS.** The experimental setup is essentially the same as that used in our previous work [13] and the scattering geometry for the x-ray scattering with a pulsed magnetic field is shown in the Extended Data Fig. 4.

The high magnetic field experiment was performed at the X-ray Correlation Spectroscopy (XCS) instrument of the Linac Coherent Light Source (LCLS) at the SLAC National Accelerator Laboratory [32]. We use the pink beam with an incident photon energy of 8.8 keV and a horizontally (sigma) polarized x-ray, just below than the Cu *K*-edge to reduce the fluorescence background from copper. Femtosecond x-ray pulses were synchronized to arrive at the sample when the magnetic field pulse reached the maximum. Note that the maximum strength of magnetic field in this work is 32.1 T, which is higher than our previous work [13].

**Data Acquisition.** All field-applied data were collected by synchronizing a magnetic pulse and x-ray pulse [13]. To obtain the zero-field background, diffraction-patterns were collected before and after the magnetic pulse. In the main text, all difference maps of diffraction-patterns are produced by subtracting the zero-field data. As an example, field-applied and zero-field diffraction intensity maps of Fig. 1c are shown in Extended Data Fig. 1.

**Additional References**

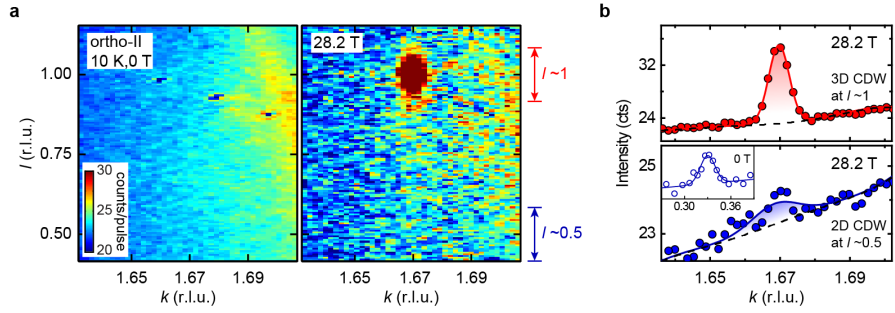

**Extended Data Figure 1| Raw diffraction pattern. a,** Diffraction intensity of ortho-II in the $kl$-plane near (0, 2-$q$, 1) at zero magnetic field (left panel) and at high magnetic field (right panel). Note that the feature around $k \sim 1.7$ and $l \sim 0.8$ is a condensation of water molecules results in the formation of an "ice ring" in the vicinity of the CDW signal [13]. **b,** Projected intensities from the diffraction pattern at $\mu_0 H$ = 28.2 T near $l \sim 1$ (upper panel) and $l \sim 0.5$ (lower panel). It indicates the coexistence of 3D and quasi-2D CDW orders in YBCO ortho-II under the high magnetic field. Inset (RSXS data) shows the quasi-2D CDW orders in YBCO ortho-II at $\mu_0 H$ = 0 T and $T \sim 20$K. Colored solid lines are Gaussian fits to the high-field data with polynomial backgrounds at zero-field (black solid line).

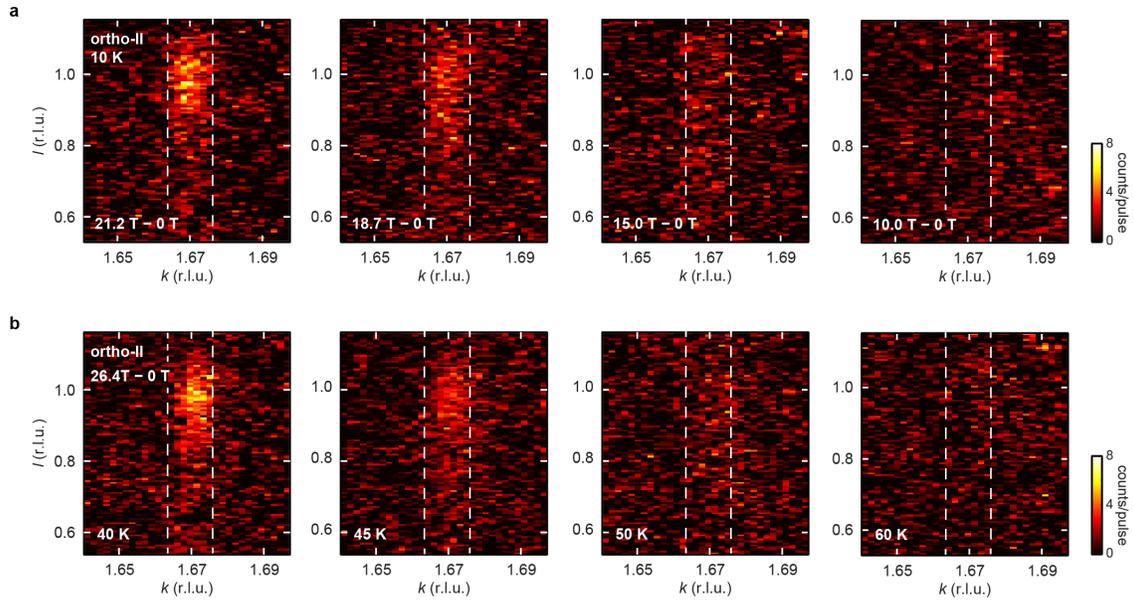

**Extended Data Figure 2| Zero-field-background-subtracted diffraction intensity maps of field induced 3D CDW of ortho-II. a,** Magnetic field dependencies. **b**, Temperature dependencies. Dashed lines indicate the windows used to deduce the projected peak profiles in Figs. 2a and 2d.

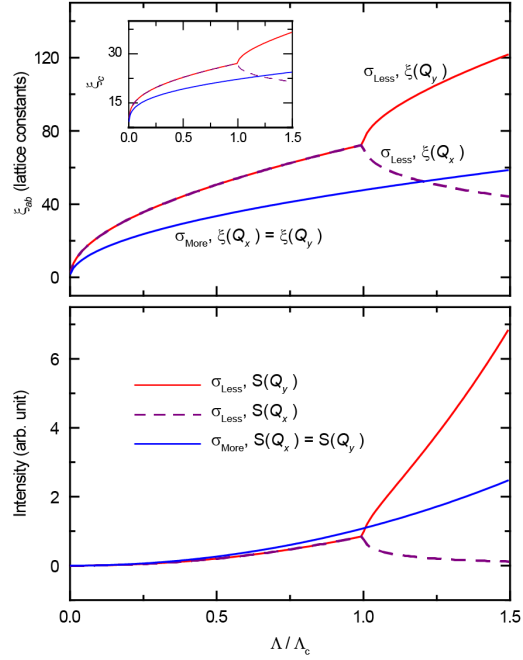

**Extended Data Figure 3| Numerically computed CDW order.** Correlation lengths (upper panel) and peak intensities (lower panel) associated with the $x$ and $y$ directed components of the CDW order parameter as a function of increasing mean squared magnitude of the local CDW order parameter for the "Less" and "More" disordered regions. We have taken $\sigma_{More} = 1.5\sigma_{Less}$, $V_z = 0.1$, and for graphical clarity have rescaled $S$ for the more disordered regions by a factor of 3. The main panel a shows the in-plane correlation lengths, while the inset shows the correlation length with the $z$-direction (*e.g.*, $c$-axis). Here, $\Lambda_c$ is the critical value of the CDW magnitude in the Less disordered regions, and for simplicity we show results for $T = 0$. $\xi_{ab}(\boldsymbol{Q})$ and $\xi_c(\boldsymbol{Q})$ are the in-plane and out-of-plane correlation lengths corresponding to ordering vector $\boldsymbol{Q}$, respectively.

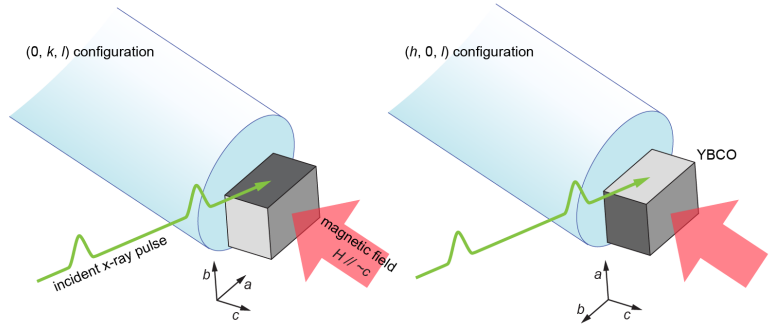

**Extended Data Figure 4| Experimental geometry.** Crystallographic orientations of YBCO crystals used for the measurement of the $(0, k, l)$ reflection (left) and $(h, 0, l)$ reflection (right). The green line represents the incident x-ray free electron pulses. The red arrow indicates the direction of the magnetic field.

# "Ideal charge density wave order in the high-field state of superconducting YBCO"

H. Jang,[1] W.-S. Lee,[2] H. Nojiri,[3] S. Matsuzawa,[3] H. Yasumura,[3] L. Nie,[4] A. V. Maharaj,[4] S. Gerber,[5] Y. Liu,[1] A. Mehata,[1] D. A. Bonn,[6,7] R. Liang,[6,7] W. N. Hardy,[6,7] C. A. Burns,[1,8] Z. Islam,[9] S. Song,[10] J. Hastings,[10] T. P. Devereaux,[2] Z.-X. Shen,[2,4] S. A. Kivelson,[4] C.-C. Kao,[1] D. Zhu,[10] and J.-S. Lee[1,*]

[1]Stanford Synchrotron Radiation Lightsource, SLAC National Accelerator Laboratory, Menlo Park, California 94025, USA
[2]Stanford Institute for Materials and Energy Science,
SLAC National Accelerator Laboratory and Stanford University, Menlo Park, California 94025, USA
[3]Institute for Materials Research, Tohoku University, Katahira 2-1-1, Sendai, 980-8577, Japan
[4]Geballe Laboratory for Advanced Materials, Departments of Physics and Applied Physics,
Stanford University, Stanford, California 94305, USA
[5]SwissFEL, Paul Scherrer Institut, 5232 Villigen PSI, Switzerland
[6]Department of Physics & Astronomy, University of British Columbia, Vancouver, British Columbia, Canada V6T 1Z1
[7]Canadian Institute for Advanced Research, Toronto, Ontario, Canada M5G 1Z8
[8]Dept. of Physics, Western Michigan University, Kalamazoo, Michigan, 49008, USA
[9]The Advanced Photon Source, Argonne National Laboratory, Argonne, Illinois 60439, USA
[10]Linac Coherent Light Source, SLAC National Accelerator Laboratory, Menlo Park, California 94025, USA
[11]SLAC National Accelerator Laboratory, Menlo Park, California 94025, USA

Our proposed interpretation of the X-ray results involves two basic assumptions: 1) In the absence of disorder, the intrinsic electronic correlations favor the formation of a unidirectional, incommensurate CDW, and 2) There are large scale inhomogeneities in the sample, which we treat schematically as regions of "More" and "Less" disorder. In this Supplement, we discuss how such assumptions can lead naturally to a sharp crossover in the directionality and correlation length of the CDW as the magnetic field is increased. Figure 5 in the main text illustrates the basic physics - in the $\sigma - \Lambda$ plane where $\sigma$ characterizes the strength of disorder and $\Lambda$ is the strength of the magnetic field (an increasing function of the magnetic field), there is a sharp crossover from short ranged bidirectional CDW correlations to longer ranged, strongly unidirectional CDW correlations. In a tetragonal cuprate, this would be associated with a thermodynamic phase transition to a nematic state, a form of "vestigial" order. We assume that at low fields both More and Less disordered regions are in the isotropic phase, while at low temperatures above a critical field strength the Less disordered regions enter the nematic phase, thus exhibiting strong unidirectional character.

**A classical effective field theory:** While many of the major points that underlie our proposal follow on rather general grounds from the statistical mechanics of disordered systems, the most straightforward way to illustrate them is by turning to the solution of the effective model of an incommensurate CDW introduced in Ref. [1]. We consider a classical effective field theory with two complex fields, $\psi_x(\vec{r})$ and $\psi_y(\vec{r})$, representing the slowly varying amplitude of a CDW at wave vectors $\vec{Q}_x = q\hat{x}$ and $\vec{Q}_y = q\hat{y}$, respectively. A biquadratic coupling of the form $2\Delta|\psi_x|^2|\psi_y|^2$ appears in the effective action, where we take $\Delta > 0$ which favors unidirectional (stripe) over bidirectional (checkerboard) order. This model can be solved in the self-consistent Gaussian approximation using the replica trick to treat the disorder. The results are controlled in a formal $N \to \infty$ limit in which $\psi_x$ is an $O(N)$ vector; it has been shown to agree qualitatively with results of Monte-Carlo simulations on the same model for the physical ($N = 2$) case in Ref. [2].

As we are only interested in qualitative results, we will simplify the problem at the expense of neglecting several material specific features of YBCO: We consider a model with one plane per unit cell and assume an interlayer coupling, $V_z > 0$, that favors in-phase interplane ordering. Moreover, we have not included the in-plane anisotropy of the CDW stiffness constants, i.e. in the notation of Ref. [1] we have taken $\kappa_\parallel = \kappa_\perp = 1$. (The final equality amounts to a particular choice of units of in-plane length, $b = 1$.)

For this simplified model, (for in plane momenta $k_x^2 + k_y^2 \ll q^2$) the CDW structure factor can be expressed as

$$S(\vec{k} + \vec{Q}_x) = TG(\vec{k}, \mu + \mathcal{N}) + \sigma^2|G(\vec{k}, \mu + \mathcal{N})|^2 \quad (1)$$
$$S(\vec{k} + \vec{Q}_y) = TG(\vec{k}, \mu - \mathcal{N}) + \sigma^2|G(\vec{k}, \mu - \mathcal{N})|^2$$

where

$$G^{-1}(\vec{k}, \mu) = \mu + k_x^2 + k_y^2 + V_z[1 - \cos(k_z c)]. \quad (2)$$

$G$ can be recognized as the simple Ornstein Zernicke form of the order parameter correlations in a generic system in the disordered phase proximate to a critical point. $k_z$ is the out-of-plane dispersion and $c$ is the $c$-axis lattice parameter. The only subtlety is that the effective chemical potential ($\mu$) and the "nematic order parameter" ($\mathcal{N}$), are determined from the self-consistency equations,

$$\Lambda(T, H) = \int \frac{d\vec{k}}{(2\pi)^3}[S(\vec{k} + \vec{Q}_x) + S(\vec{k} + \vec{Q}_y)] \quad (3)$$

where $\Lambda$ is the mean squared amplitude of the CDW order parameter (assumed to be an otherwise known function of $T$ and $H$), and

$$\mathcal{N} = \mathcal{N}_0 + 2\Delta \int \frac{d\vec{k}}{(2\pi)^3} [S(\vec{k} + \vec{Q}_y) - S(\vec{k} + \vec{Q}_x)] \quad (4)$$

where $\mathcal{N}_0$ is an intrinsic nematicity (due to the orthorhombicity of the crystal). Again, for simplicity, we will henceforth report results in the limit $\mathcal{N}_0 \to 0^+$. In this case, $\mathcal{N} = 0$ is always a possible solution of the self-consistency equations. However, if the CDW ordering tendency is sufficiently strong (i.e. for $\Lambda(T, H) > \Lambda_c$ where the critical value $\Lambda_c$ is itself a function of $T$ and $\sigma$), a second solution with $\mathcal{N} > 0$ is preferred; this is the nematic phase which spontaneously breaks the point-group symmetry. Consistent with general theorems, so long as $\sigma > 0$, the CDW order inferred from these equations always has a finite correlation length, i.e. *there is never long-range CDW order* in the presence of even weak quenched randomness.

This set of equations was analyzed under various circumstances in Ref. [1]. Generally, as the system is tuned from the isotropic to the nematic phase in any fashion (for instance, by increasing $\Lambda$) several changes in the nature of the correlations onset very rapidly at the transition point: The in-plane correlation length, $\xi_{ab}$, grows substantially as does the interplane correlation length, $\xi_c$, so long as $V_z$ is not too small. The degree of directionality, $S(\vec{Q}_y)/S(\vec{Q}_x)$, which is 1 in the isotropic phase, becomes rapidly much larger than 1 in the nematic phase. Simultaneously, the peak intensity, $S(\vec{Q}_y)$, shows the most dramatic increase of all.

As an illustrative example, we consider the simplest case of $T = 0$ and vanishingly small interplane [3] coupling, $V_z \to 0$. In this limit, the various integrals can be evaluated analytically. The self-consistency equations (which one is to solve for $\mu$ and $\mathcal{N}$) become

$$\Lambda(0, H) = \left(\frac{\sigma^2}{2\pi}\right)\left[\frac{\mu}{\mu^2 - \mathcal{N}^2}\right]$$
$$\mathcal{N} = 2\Delta \left(\frac{\sigma^2}{2\pi}\right)\left[\frac{\mathcal{N}}{\mu^2 - \mathcal{N}^2}\right] \quad (5)$$

from which it follows that the critical value of $\Lambda$ for the occurrence of the nematic phase is

$$\Lambda_c = \frac{\sigma}{2\sqrt{\pi}\Delta} . \quad (6)$$

For $\Lambda < \Lambda_c$, $\mu = \sigma^2/2\pi\Lambda$ and the in-plane correlation lengths and intensities of the peaks at $\vec{Q}_x$ and $\vec{Q}_y$ are

$$\xi_{ab}(\vec{Q}_x) = \xi_{ab}(\vec{Q}_y) = \frac{1}{\sqrt{\mu}} \quad \text{and}$$
$$S(\vec{Q}_x) = S(\vec{Q}_y) = 2\pi\Lambda |\xi_{ab}(\vec{Q}_y)|^4 . \quad (7)$$

For $\Lambda > \Lambda_c$, $\mu = 2\Delta\mathcal{N}$, $\mathcal{N} = 2\Delta\sqrt{[\Lambda^2 - \Lambda_c^2]}$, and

$$\xi_{ab}(\vec{Q}_x) = \frac{1}{\sqrt{\mu + \mathcal{N}}} , \quad \xi_{ab}(\vec{Q}_y) = \frac{1}{\sqrt{\mu - \mathcal{N}}} , \quad (8)$$

and

$$\frac{S(\vec{Q}_y)}{S(\vec{Q}_x)} = \left[\frac{\xi_{ab}(\vec{Q}_y)}{\xi_{ab}(\vec{Q}_x)}\right]^4 . \quad (9)$$

To address the growth of 3D correlations in the nematic phase, it is obviously necessary to include explicitly the effects of non-zero $V_z$. While the self consistency equations are still analytically tractable, the solutions are sufficiently complicated that we only evaluate them numerically. The inter-plane correlation length can be computed as

$$\xi_c = (c/2)\left[\text{arcsinh}\left(1/\xi_{ab}\sqrt{2V_z}\right)\right]^{-1}. \quad (10)$$

Clearly, there is strong tendency to increased 3D order when the in-plane correlations become sufficiently long.

In the extended data Fig. 3, we show the evolution of these quantities computed numerically from the full self-consistency equations with $V_z = 0.1$ as a function of $\Lambda$ for two different values of $\sigma$, one representative of the More and one of the Less disordered regimes, under the assumption $\sigma_{\text{More}}/\sigma_{\text{Less}} = 1.5$. To make contact with experiment, one should imagine that the field dependence of $\Lambda$ determined by competition with superconductivity, $\Lambda = \Lambda_0[1 - |\phi|^2]$, where the amplitude of the superconducting order $\phi$ is (presumably) a decreasing function of increasing $H$. There is clearly a sharp increase in the in-plane correlation lengths and peak intensities at the nematic phase transition.

Finally, it is tempting to fit the observed X-ray scattering intensities to our theoretical model in order to make estimates of the volume fractions of the More and Less disordered regions from relative strengths of the various contributions to the X-ray scattering intensity, $I(\vec{k})$. However, making connection between the calculated and measured quantities carries with it additional uncertainties. $I(\vec{k})$ is generally dominated by scattering off the atomic cores, and so is a measure of the atomic displacements. A periodic pattern of atomic displacements proportional to the CDW order parameter, $\psi$, is generic. However, while positions and widths of the observable peaks in $I(\vec{k})$ reflect the pattern of translation symmetry breaking and the CDW correlation length, the variation of intensity from one Brillouin zone to the other encodes the precise pattern of induced lattice distortion. Crudely,

$$I(\vec{k} + \vec{K}) \approx I_0(\vec{k})S(\vec{k}) \quad (11)$$

where $\vec{K}$ is a reciprocal lattice vector, and $I_0(K)$ is a complicated structure factor. One would generally expect substantial scattering intensity in a very large number of Brillouin zones, at least hundreds. If the scattering intensities were measured in sufficiently many different Brillouin zones, one could work backwards to the actual lattice displacements induced by the CDW. This has been undertaken for the low field diffraction data in Ref. [4].

Because of the constraints of the high field experiment, our data is limited to little over one Brillouin zone, so such a refinement is not possible. Indeed, because we cannot study the scattering intensity in multiple Brillouin zones, and so have little information about the $\vec{K}$, $T$, and $H$ dependences of $I_0$, we are reluctant to interpret the intensities of the various features of the X-ray data. Rather we focus on features associated with the "shape" of the diffraction peak, from which the correlation lengths can be obtained, and $S(\vec{Q}_x)/S(\vec{Q}_y)$.

---

* jslee@slac.stanford.edu